\documentclass[lettersize,journal]{IEEEtran}
\usepackage{amsmath,amsfonts}
\usepackage{algorithmic}
\usepackage{algorithm}
\usepackage{array}
\usepackage[caption=false,font=normalsize,labelfont=sf,textfont=sf]{subfig}
\usepackage{textcomp}
\usepackage{stfloats}
\usepackage{url}
\usepackage{hyperref}
\usepackage{verbatim}
\usepackage{graphicx}
\usepackage{cite}
\usepackage{orcidlink}
\usepackage{tikz}
\usepackage{multirow}
\usetikzlibrary{positioning}
\usepackage{threeparttable}
\usepackage{booktabs}
\usepackage{fancyhdr}

\begin{document}

\title{Fooling the Mind Twice: Human Detection of Audiovisual Deepfakes Under Repeated Exposure and Corrective Feedback} 

\author{%
Ammarah Hashmi $^{*}$\,\orcidlink{0000-0002-1973-6902},~%
Sahibzada Adil Shahzad $^{*}$\,\orcidlink{0009-0000-5591-8423},~%
Yu Tsao\,\orcidlink{0000-0001-6956-0418},~\IEEEmembership{Senior Member,~IEEE},~%
and \\ Hsin-Min Wang\,\orcidlink{0000-0003-3599-5071},~\IEEEmembership{Senior Member,~IEEE}

\thanks{Ammarah Hashmi is with the Institute of Information Science, Academia Sinica, Taipei 11529, Taiwan. (e-mail: hashmiammarah0@gmail.com).

Sahibzada Adil Shahzad is with the Institute of Information Science, Academia Sinica, Taipei 11529, Taiwan. (e-mail: adilshah275@iis.sinica.edu.tw).

Yu Tsao is with the Research Center for Information Technology Innovation, Academia Sinica, Taipei 11529, Taiwan. (e-mail:
yu.tsao@citi.sinica.edu.tw).

Hsin-Min Wang is with the Institute of Information Science, Academia Sinica, Taipei 11529, Taiwan. (e-mail: whm@iis.sinica.edu.tw).}
\thanks{\hrulefill \newline$^{*}$ Equal contribution. Author order is alphabetical.}}

\maketitle
\thispagestyle{fancy} 
\begin{abstract}

Human deepfake detection remains unreliable, and whether reassessment with corrective feedback improves subsequent judgments remains unclear. This study examined differences between initial and repeated deepfake assessments among 110 participants using audiovisual stimuli from the FakeAVCeleb and Deepfake Detection Challenge (DFDC) datasets. Participants evaluated 40 videos from each dataset in phase 1 (PhaseI) and reassessed the same videos in phase 2 (PhaseII) under a procedure combining repeated presentation and corrective feedback. For the FakeAVCeleb dataset, the participants' mean accuracy increased by 4.68 percentage points from 63.30\% in PhaseI to 67.98\% in PhaseII ($t(109)=5.61$, $p<.001$, $d_z=0.54$). A judgment-level logistic regression with standard errors clustered by participant and video produced a consistent result: $\beta=0.208$, $SE=0.040$, $z=5.23$, $p<.001$, $\mathrm{OR}=1.23$. For the DFDC dataset, the participants' mean accuracy increased by 8.64 percentage points from 59.52\% in PhaseI to 68.16\% in PhaseII ($t(109)=7.37$, $p<.001$, $d_z=0.70$); the logistic model yielded: $\beta=0.375$, $SE=0.065$, $z=5.78$, $p<.001$, $\mathrm{OR}=1.46$. Improvements primarily came from fewer false negatives, while false-positive rates remained unchanged. However, judgment reversals occurred in 27.55\% of FakeAVCeleb and 32.91\% of DFDC assessments, with some initially correct judgments becoming incorrect upon reassessment. Despite the accuracy gains, repeated assessment did not eliminate judgment instability or residual error. Participants also exhibited consistent overconfidence, with mean self-rated confidence exceeding actual accuracy by 11.99 percentage points for FakeAVCeleb and 12.82 percentage points for DFDC across both phases. These findings indicate that repetition with corrective feedback moderately improves deepfake detection but does not fully resolve the limitations of human judgment.

\end{abstract}

\begin{IEEEkeywords}
audiovisual deepfake detection, human perception, corrective feedback, repeated assessment
\end{IEEEkeywords}

\section{Introduction}
\IEEEPARstart{D}{eepfake} technology is increasingly capable of generating or altering a person's appearance, voice, facial expressions, and audiovisual synchronization~\cite{R3}. Although this technology has beneficial applications in entertainment, accessibility, education, and creative production~\cite{R31}, it also carries risks of misinformation, fraud, defamation, and a decline in trust in digital evidence~\cite{R1, R2, R35}. The challenges posed by audiovisual deepfakes are particularly severe because observers must integrate what they see and hear to determine whether audio and video streams are authentic, manipulated, or mismatched~\cite{R3}. Similar multimodal integration challenges have been widely investigated in other domains~\cite{R36, R37}.

Most technical research has focused on automatic detection models that identify low-level artifacts, lip-sync inconsistencies, cross-modal mismatches, or manipulation traces~\cite {R6, R7, R8}.
In contrast, human-centered studies show that people often perform only slightly above chance in detecting synthetic media, frequently confuse real and fake content, and overestimate their abilities~\cite{R10, R11, R12, R13}.
This highlights a gap between machine detection accuracy and human perceptual reliability~\cite{R34}. 
However, existing studies primarily examine first-exposure judgments, leaving the effects of repeated encounters with manipulated content largely unexplored.

In real-world media environments, people may encounter the same or similar audiovisual content repeatedly through social media, news, messaging platforms, and discussions~\cite{R26}. Repetition may improve detection by enhancing cue recognition and memory, but it may also increase familiarity, reinforce initial judgments, or enhance perceived credibility~\cite{R14, R25, R33}. These opposing effects relate to cognitive mechanisms such as perceptual learning, memory, familiarity, the mere-exposure effect, the illusory-truth effect, and source monitoring~\cite{R19, R20, R21, R22}. Therefore, understanding human vulnerability to audiovisual deepfakes requires examining how detection judgments change across repeated exposures rather than focusing only on initial encounters.

This paper examines human perception of audiovisual deepfakes through a two-phase repeated-exposure design with corrective feedback after the initial judgments. Beyond first-exposure detection accuracy, we examine how judgments change when participants re-evaluate the same content after receiving ground-truth information. We address the following research questions:

\begin{itemize}
    \item \textbf{RQ1:} How accurately do participants distinguish real from fake audiovisual videos, and how does their accuracy change from phase 1 (PhaseI) to phase 2 (PhaseII) after repeated exposure and corrective feedback?
    \item \textbf{RQ2:} To what extent do repeated judgments exhibit decision consistency, judgment reversal, correct persistence, error persistence, error correction, and new error formation, and how do false-positive (misclassified as fake) and false-negative (misclassified as real) rates change from PhaseI to PhaseII?
    \item \textbf{RQ3:} To what extent are changes in detection accuracy, error rates, and judgment-transition patterns consistent between the two phases in the FakeAVCeleb and DFDC datasets? 
\end{itemize}

To answer these questions, we conducted a two-phase within-subjects study in which the same participants evaluated the same audiovisual stimuli twice, with presentation order randomized independently for each participant and phase. In the primary study, 110 participants evaluated 20 real and 20 fake videos from the FakeAVCeleb dataset~\cite{R1} in PhaseI and re-evaluated the same videos in PhaseII. After each judgment, participants received binary corrective feedback indicating the ground-truth authenticity label. At the end of the study, each participant was asked to provide an overall self-rated confidence level. This design produced 8,800 responses, which are used to analyze changes in accuracy, decision consistency, judgment reversals, correctness transitions, false-positive rates, false-negative rates, and participant-level variability across phases. We also applied the two-phase protocol to stimuli from the DFDC dataset~\cite{R2} to examine whether these cross-phase behavioral patterns were generalized across the datasets. The main contributions of this work are as follows:
\begin{itemize}
    \item We conducted a two-phase within-subjects study, collecting 8,800 responses from 110 participants on the FakeAVCeleb dataset to examine changes in audiovisual-deepfake judgments after repeated exposure and corrective feedback. We replicated the protocol using stimuli from the DFDC dataset for cross-dataset validation.
    \item We examine audiovisual-deepfake perception as a repeated decision-making process, extending the analysis beyond one-shot detection accuracy.
    \item We show that mean detection accuracy improves from PhaseI to PhaseII, but the improvement is practically modest and persistent errors remain.
    \item We operationalize repeated-judgment behavior using decision consistency, judgment reversal, correct persistence, error persistence, error correction, new error formation, false-positive rates, and false-negative rates.

\end{itemize}

The remainder of this paper is organized as follows. Section~\ref{sec:related_work} reviews related work. Section~\ref{sec:experimental_design} describes the experimental design, datasets, repeated-exposure procedure, and validation study. 
Section~\ref{sec:results} presents the results, and Section~\ref{sec:discussion} discusses the findings. Sections~\ref{sec:limitations} and~\ref{sec:future_work} summarize the limitations and future directions, respectively. Finally, Section~\ref{sec:conclusion} concludes the paper.

\section{Related Work}
\label{sec:related_work}

\subsection{Human Perception of Synthetic Media}

Human detection of manipulated media has been studied in images~\cite{R16}, audio~\cite{R15}, video~\cite{R10}, and multimodal content~\cite{R12}. Research on synthetic images has shown that observers are often deceived by manipulated or AI-generated faces and may even judge synthetic faces as trustworthy \cite{R16}, \cite{R17}, \cite{R18}. Studies of audio deepfakes have also found that listeners cannot reliably identify synthetic speech, even when the task is explicitly framed as deception detection~\cite{R13, R15}. For video-based manipulations, human observers frequently confuse real and fake content, overestimate their detection ability, and rely on inconsistent or nondiagnostic visual cues~\cite{R10, R11, R12}.

Audiovisual deepfakes are challenging to evaluate because detection requires integrating auditory and visual information, while manipulations may affect either or both modalities without disrupting perceptual coherence~\cite{R1, R3}. Although humans can detect some manipulated media, their performance is considerably lower than specialized detection models~\cite{R23, R12}. Similarly, large language models can achieve human-level performance with forensic prompting but remain below dedicated audiovisual detectors~\cite{R20}. These findings highlight human vulnerability to audiovisual manipulation; however, existing studies largely focus on first-exposure judgments and overlook how decisions change after corrective feedback and repeated evaluation.

\subsection{Automatic Audiovisual Deepfake Detection}

Automatic deepfake detection methods use visual artifacts, speech-lip inconsistencies, temporal irregularities, acoustic traces, or multimodal feature fusion~\cite{R4, R5, R6, R24}. Visual approaches identify facial artifacts\cite{R27}, unnatural motion~\cite{R28}, or generation inconsistencies~\cite{R30}. Audio-based approaches analyze acoustic evidence of synthesis or manipulation~\cite{R29}, while cross-modal approaches model correspondences between speech and mouth movements~\cite{R9}. More recent systems combine auditory and visual representations to detect inconsistencies that may not be evident from either modality alone~\cite{R6, R8, R24}.

In many controlled benchmark settings, automatic detectors often outperform human observers. Using the same FakeAVCeleb stimuli, previous research reported detector accuracies of 87.50\% to 97.50\%, compared with mean human accuracy of 63.30\%~\cite{R23}. This gap suggests that automated systems capture subtle artifacts and cross-modal inconsistencies that humans may overlook. However, detector benchmarks do not reveal how human judgments change with repetition, feedback, or familiarity, which are critical factors in real-world authenticity assessment and human-machine interaction.

\subsection{Memory, Familiarity, and Repeated Exposure}

Cognitive psychology provides several mechanisms relevant to repeated encounters with deepfake content. The mere-exposure effect suggests that repetition can increase familiarity and affective preference~\cite{R19}. The illusory-truth effect shows that repeated statements may be judged more truthful, even when they are false~\cite{R20, R21}. Source-monitoring theory explains how people may remember content while misattributing its source, origin, or truth status~\cite{R22}. In the context of audiovisual deepfakes, viewers may recall having seen a video without accurately recalling whether it was authentic, manipulated, or previously identified as fake.

Repeated exposure may therefore produce contradictory effects. On the one hand, it may facilitate learning by drawing the observer's attention to manipulation cues, retrieving previous corrective feedback, and revising earlier judgments. On the other hand, repetition may also reinforce familiarity, retain initial errors, or increase confidence in erroneous conclusions~\cite{R25}. The final outcome depends on whether the observer uses the second exposure to retrieve and apply diagnostic information or to interpret familiarity as evidence of authenticity. Thus, repeatedly viewed deepfakes may be easier for some observers to identify, while they may be more convincing for others.

Most research on human deepfake perception has emphasized the first-exposure detection accuracy, warning effects, or comparisons between human and machine performance~\cite{R10, R11, R12, R13, R15, R16, R17, R23}. Comparatively little is known about how people respond when they evaluate the same audiovisual stimuli again after receiving corrective feedback. Prior work has not adequately characterized whether initial errors persist, whether incorrect judgments are corrected, whether initially correct judgments become incorrect, or whether these transition patterns generalize across datasets. The present study addresses these gaps through a two-phase repeated-judgment design that examines the cross-phase accuracy change, judgment stability, error persistence, error correction, new error formation, and consistency of these behavioral patterns across datasets.

\section{Experimental Design}
\label{sec:experimental_design}
\subsection{Participants}
A total of 110 English-speaking participants completed the repeated-exposure experiment. Participants were required to self-report normal or corrected-to-normal vision and normal hearing. The final sample was balanced in terms of gender and native-language background, including 55 male and 55 female participants, with 55 native and 55 non-native English speakers. The participants were roughly evenly distributed in three age groups: 20-30 years ($n=37$), 31-40 years ($n=37$), and 41-50 years ($n=36$). Self-reported information technology (IT) proficiency was recorded at four levels: beginner ($n=28$), intermediate ($n=28$), good ($n=27$), and expert ($n=27$). Participants who did not complete all rounds were excluded from the analysis. The basic demographic information of the participants is summarized in Table~\ref{tab:participant_demographics}. 

\begin{figure*}[!t]
    \centering
    \includegraphics[width=0.90\textwidth]{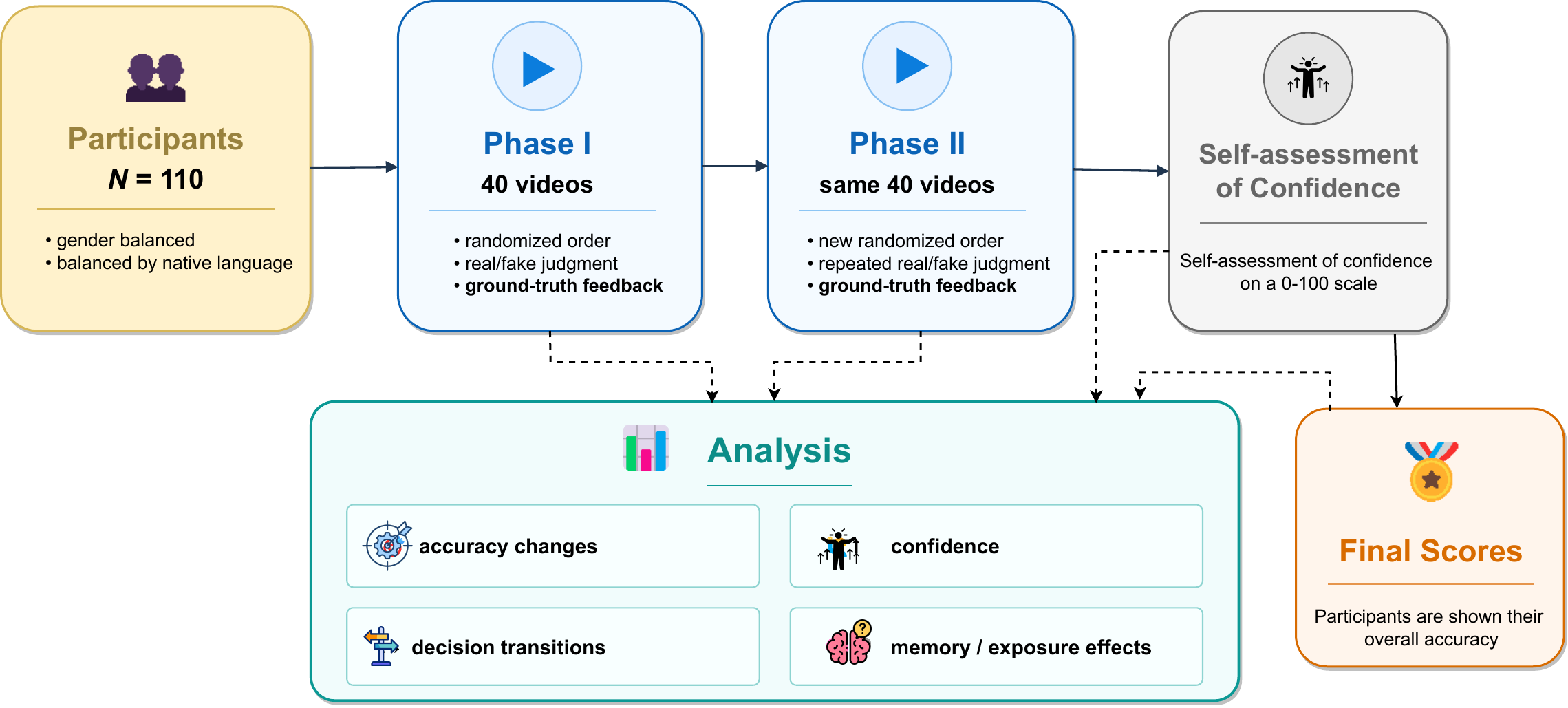}
    \caption{The two-phase repeated-exposure experimental workflow.}
    \label{fig:ExperimentDesign}
\end{figure*}

\begin{table}[t]
\centering
\caption{Participant demographics for the repeated-exposure with corrective feedback experiment.}
\label{tab:participant_demographics}
\begin{tabular}{l|l|c|c}
\hline 
\textbf{\begin{tabular}{@{}c@{}}Demographic \\ variable\end{tabular} } & \textbf{Category} & \textbf{n} & \textbf{Percentage (\%)} \\
\hline\hline
Total sample & \begin{tabular}{@{}l@{}}English-speaking \\ participants\end{tabular}  & 110 & 100.0 \\
\hline
\multirow{2}{*}{Gender} & Male & 55 & 50.0 \\ \cline{2-4}
       & Female & 55 & 50.0 \\
\hline
\multirow{2}{*}{\begin{tabular}[c]{@{}l@{}}Native-language \\ background\end{tabular}}  & \begin{tabular}{@{}l@{}}Native English \\ speakers\end{tabular}    & 55 & 50.0 \\ \cline{2-4}
                           & \begin{tabular}{@{}l@{}}Non-native English\\ speakers\end{tabular}  & 55 & 50.0 \\
\hline
\multirow{3}{*}{Age group} & 20-30 years & 37 & 33.6 \\ \cline{2-4}
          & 31-40 years & 37 & 33.6 \\ \cline{2-4}
          & 41-50 years & 36 & 32.7 \\
\hline

\hline
\multirow{4}{*}{\begin{tabular}[c]{@{}l@{}}Self-reported \\ IT proficiency\end{tabular}} 
& Beginner     & 28 & 25.5 \\ \cline{2-4}
& Intermediate & 28 & 25.5 \\ \cline{2-4}
& Good         & 27 & 24.5 \\ \cline{2-4}
& Expert       & 27 & 24.5 \\
\hline
\end{tabular}
\vspace{-4mm}

\end{table}

\subsection{Datasets and Stimuli}
The primary experiment used 40 audiovisual stimuli selected from the FakeAVCeleb dataset~\cite{R1}. The stimulus set comprised 20 real and 20 fake videos and was balanced in terms of speaker gender. The fake subset included audio-only, visual-only, and audiovisual manipulations, providing representation of different forms of synthetic media alteration.

For cross-dataset validation, we selected 40 frontal face videos from the DFDC dataset~\cite{R2}, comprising 20 real and 20 fake videos. The DFDC stimuli exhibited more varied recording conditions than the FakeAVCeleb stimuli, including variation in illumination, background, pose, facial expression, and video quality. 

\subsection{Two-Phase Repeated-Exposure Procedure}

Participants performed a binary authenticity judgment task in which they classified each video as real or fake. In PhaseI, each participant viewed all 40 videos in randomized order and provided a real/fake judgment for each stimulus. When a video was judged to be fake, the participant additionally indicated the perceived source of manipulation by selecting audio, visual, or both. After each judgment, the interface displayed binary corrective feedback indicating the ground-truth real/fake label. 

In PhaseII, participants evaluated the same 40 videos in a newly randomized order. The presentation order was randomized independently for each participant and phase. This design enabled the measurement of cross-phase changes in detection accuracy, response stability, error persistence, error correction, and new error formation after prior exposure and corrective feedback. After completing both phases, participants were asked to self-rate their overall confidence in their judgments. The experimental workflow is shown in Fig.~\ref{fig:ExperimentDesign}.

\subsection{Web Platform}

The experiment was administered through a web-based interface that presented audiovisual stimuli, collected participant responses, delivered binary corrective feedback, and recorded confidence ratings. Each participant completed the task individually using their own device and was instructed to watch each video with audio enabled before submitting a response.

For each round, the platform recorded the participant ID, stimulus ID, phase, randomized presentation order, binary real/fake judgment, perceived manipulation modality when applicable, and response correctness. The video order was independently randomized for each participant and phase. At the end of the study, participants provided an overall confidence score reflecting their confidence in the judgments made during the experiment. The platform therefore ensured consistency across all participants in stimulus presentation, response logging, feedback delivery, repeated-exposure tracking, and confidence score collection.

\subsection{Ethics, Consent, and Data Handling}

Participants were recruited through Amazon Mechanical Turk and compensated according to the platform’s payment procedure. No personally identifiable information was used in the analysis. Each participant was assigned a unique participant ID, which was used to link PhaseI and PhaseII responses while preserving anonymity. The data collected included demographic information, self-reported IT proficiency, round-level authenticity judgments, perceived manipulation modality responses, correctness labels, and confidence ratings. 

\subsection{Repeated-Assessment and Error Metrics}
\label{subsec:error_metrics}

In this study, fake videos were treated as the positive class. Let $i\in\{1,\ldots,N\}$ denote the participant, $j\in\{1,\ldots,J\}$ denote the video, and $p\in\{1,2\}$ denote the assessment phase, where $N=110$ and $J=40$ for each dataset. Each dataset contained 20 real and 20 fake videos. Let $g_j\in\{0,1\}$ denote the ground-truth label of video $j$, where 0 represents real and 1 represents fake, and 
$y_{ij}^{(p)}\in\{0,1\}$ where 1 indicates that the judgment is false and 0 indicates that the judgment is true.


The accuracy for participant $i$ in phase $p$ was calculated as
\begin{equation}
A_i^{(p)}
=
\frac{1}{J}
\sum_{j=1}^{J}
\mathbb{I}
\left(
y_{ij}^{(p)}=g_j
\right)
\times 100\%,
\label{eq:accuracy}
\end{equation}
where $\mathbb{I}(\cdot)$ denotes the indicator function. The cross-phase accuracy difference for participant $i$ was defined as
\begin{equation}
\Delta A_i
=
A_i^{(2)}-A_i^{(1)}.
\label{eq:accuracy_change}
\end{equation}
Since fake videos were treated as the positive class, the false positive error corresponded to a real video classified as fake, whereas the false-negative error corresponded to a fake video classified as real. The false-positive rate (FPR) for participant $i$ in phase $p$ was calculated as
\begin{equation}
\mathrm{FPR}_i^{(p)}
=
\frac{
\displaystyle
\sum_{j:g_j=0}
\mathbb{I}
\left(
y_{ij}^{(p)}=1
\right)
}{
J_{\mathrm{Real}}
}
\times 100\%,
\label{eq:fpr}
\end{equation}
and the false-negative rate (FNR) for participant $i$ in phase $p$ was calculated as
\begin{equation}
\mathrm{FNR}_i^{(p)}
=
\frac{
\displaystyle
\sum_{j:g_j=1}
\mathbb{I}
\left(
y_{ij}^{(p)}=0
\right)
}{
J_{\mathrm{Fake}}
}
\times 100\%,
\label{eq:fnr}
\end{equation}
where $J_{\mathrm{Real}}=J_{\mathrm{Fake}}=20$.
The cross-phase difference in FPR and FNR for participant $i$ was calculated in the same way as the cross-phase accuracy difference defined in Eq. (\ref{eq:accuracy_change}); negative values indicate a reduction in the corresponding error rate.

To assess judgment stability, the judgments made by the same participant for the same video in both phases were grouped into a participant-video pair, and each pair was assigned to one of four mutually exclusive correctness-transition categories:
\begin{itemize}
    \item \emph{correct persistence}:
    $y_{ij}^{(1)}=g_j \land y_{ij}^{(2)}=g_j$;
    \item \emph{error persistence}:
    $y_{ij}^{(1)} \neq g_j \land y_{ij}^{(2)} \neq g_j$;
    \item \emph{error correction}:
    $y_{ij}^{(1)} \neq g_j \land y_{ij}^{(2)}=g_j$; and
    \item \emph{new-error formation}:
    $y_{ij}^{(1)}=g_j \land y_{ij}^{(2)} \neq g_j$.
\end{itemize}

The cross-phase response consistency rate was defined as
\begin{equation}
C
=
\frac{1}{NJ}
\sum_{i=1}^{N}
\sum_{j=1}^{J}
\mathbb{I}
\left(
y_{ij}^{(1)}=y_{ij}^{(2)}
\right)
\times 100\%.
\label{eq:consistency}
\end{equation}
The cross-phase response reversal rate was defined as
\begin{equation}
R
=
\frac{1}{NJ}
\sum_{i=1}^{N}
\sum_{j=1}^{J}
\mathbb{I}
\left(
y_{ij}^{(1)}\neq y_{ij}^{(2)}
\right)
\times 100\%.
\label{eq:reversal}
\end{equation}
Consequently, $R=100\%-C$.

The cross-phase error correction rate was defined as
\begin{equation}
E_{\mathrm{corr}}
=
\frac{
\displaystyle
\sum_{i=1}^{N}
\sum_{j=1}^{J}
\mathbb{I}
\left(
y_{ij}^{(1)} \neq g_j \land y_{ij}^{(2)}=g_j
\right)
}{
\displaystyle
\sum_{i=1}^{N}
\sum_{j=1}^{J}
\mathbb{I}
\left(
y_{ij}^{(1)} \neq g_j
\right)
}
\times 100\%.
\label{eq:error_correction}
\end{equation}
The cross-phase new-error formation rate was defined as
\begin{equation}
E_{\mathrm{new}}
=
\frac{
\displaystyle
\sum_{i=1}^{N}
\sum_{j=1}^{J}
\mathbb{I}
\left(
y_{ij}^{(1)}=g_j
\land
y_{ij}^{(2)} \neq g_j
\right)
}{
\displaystyle
\sum_{i=1}^{N}
\sum_{j=1}^{J}
\mathbb{I}
\left(
y_{ij}^{(1)}=g_j
\right)
}
\times 100\%.
\label{eq:new_error}
\end{equation}

\begin{table*}[t]
\centering
\caption{Research questions, analytical expectations, and statistical analyses.}
\label{tab:research_questions}
\renewcommand{\arraystretch}{1.15}
\setlength{\tabcolsep}{5pt}
\begin{tabular}{p{0.22\textwidth}p{0.31\textwidth}p{0.39\textwidth}}
\hline
\textbf{Research question} &
\textbf{Analytical expectation} &
\textbf{Statistical analysis} \\
\hline

RQ1: Detection accuracy and cross-phase difference &
Detection accuracy may differ between PhaseI and PhaseII under the combined repeated-exposure and corrective-feedback procedure. &
Phase-specific participant-level accuracy; paired-sample $t$ test; mean paired difference with 95\% confidence interval; within-participant effect size $d_z$; and complementary logistic regression of correctness on phase with standard errors clustered by participant and video. \\

RQ2: Judgment stability, correctness transitions, and error types &
Repeated judgments may remain stable or reverse. Initial errors may be corrected or persist, whereas initially correct judgments may remain correct or become incorrect. Cross-phase differences may vary between false-positive and false-negative errors. &
Binary-response consistency and reversal rates; binary-response and correctness-transition matrices; frequencies and percentages of correct persistence, error persistence, error correction, and new-error formation; supplementary continuity-corrected McNemar test; and paired comparisons of participant-level FPR and FNR with 95\% confidence intervals and $d_z$. \\

RQ3: Cross-dataset consistency &
The direction of PhaseI-to-PhaseII differences may be similar across FakeAVCeleb and DFDC, although their magnitudes may differ. &
Dataset-specific application of the RQ1 and RQ2 analyses; descriptive comparison of accuracy, FPR, FNR, and transition patterns; and matched within-participant comparison of dataset-specific accuracy-difference scores. \\

\hline
\end{tabular}
\end{table*}

\subsection{Confidence-Accuracy Measures}
\label{subsec:confidence_measures}

After completing both phases of the evaluation for each dataset, participants reported their overall confidence in their judgments on a scale from 0\% to 100\%. Since confidence was collected once rather than separately for each phase, it was compared with the pooled accuracy for both phases. For participant $i$, the pooled accuracy was defined as
\begin{equation}
A_{i}^{\mathrm{pool}}
=
\frac{A_i^{(1)}+A_i^{(2)}}{2}.
\label{eq:pooled_accuracy}
\end{equation}
The signed confidence-accuracy gap for participant $i$ was calculated as
\begin{equation}
G_{i}
=
C_{i}-A_{i}^{\mathrm{pool}}.
\label{eq:signed_confidence_gap}
\end{equation}
A positive value indicates that the participant's self-rated confidence exceeds the observed accuracy, i.e., overconfidence; conversely, a negative value indicates that confidence is lower than actual performance.
The absolute confidence-accuracy gap for participant $i$ was defined as
\begin{equation}
G_{i}^{\mathrm{abs}}
=
\left|
C_{i}-A_{i}^{\mathrm{pool}}
\right|.
\label{eq:absolute_confidence_gap}
\end{equation}
This measure represents the magnitude of the confidence-accuracy discrepancy irrespective of its direction.

The signed and absolute gaps were used to characterize participant-level correspondence between self-rated confidence and observed performance. Because the confidence rating was obtained only once, at the end of both phases, these measures do not quantify phase-specific confidence changes or round-level calibration.

\subsection{Statistical Analysis Framework}
\label{subsec:statistical_analysis}

\subsubsection{General Analysis Settings}
Table~\ref{tab:research_questions} summarizes the analytical expectations and statistical analysis methods used for the three research questions. All statistical tests used in this study were two-sided, with a significance level set at $\alpha=.05$. Unless otherwise stated, analyses were conducted on FakeAVCeleb and DFDC separately. Accuracy comparisons were used as the primary analysis. The class-specific FPR and FNR comparisons, transition analyses, and confidence analyses were used as secondary analyses. 

\subsubsection{Cross-Phase Accuracy Comparison} 

To answer RQ1, accuracy was calculated separately for each participant and phase. The participant-level accuracy values in PhaseI and PhaseII were compared using a paired-sample $t$ test. We report the mean $\overline{\Delta}$ 
and standard deviation $SD_{\Delta}$
of the differences in accuracy between phases on $110$ participants $\{\Delta A_i\}_{i=1}^{110}$, 95\% confidence interval, $t$ statistic, $p$ value, and standardized within-participant effect size, $d_z$, which was calculated as
\begin{equation}
d_z = \frac{\overline{\Delta}}{SD_{\Delta}}.
\end{equation}

As a complementary judgment-level analysis, we fitted a binary logistic regression model in which response correctness was the dependent variable and experimental phase was the independent variable, with PhaseI as the reference category. Because judgments were crossed by participants and videos, we estimated two-way cluster-robust (sandwich) standard errors (clustered by both participant and video) to account for data non-independence. Wald tests were conducted using the standard normal reference distribution. We report the phase coefficient ($\beta$), two-way cluster-robust standard error, $z$ statistic, $p$ value, and odds ratio with its 95\% confidence interval. 

\subsubsection{Cross-Phase Response Transitions and Error Rates}
To answer RQ2, based on 4,400 participant-video pairs in each dataset, we compared the cross-phase response consistency rate (Eq.~(\ref{eq:consistency})), the cross-phase response reversal rate (Eq.~(\ref{eq:reversal})), the cross-phase error correction rate (Eq.~(\ref{eq:error_correction})), and the cross-phase new-error formation rate (Eq.~(\ref{eq:new_error})).

As a supplementary comparison, we used a continuity-corrected McNemar test to analyze the relative rates of cross-phase error correction and new-error formation.
However, because the participant-video pairs were clustered within both participants and videos, they violated the independence assumption of the standard McNemar test. Consequently, the test results were interpreted as supplementary to the descriptive transition counts and percentages, rather than as definitive inferential evidence for judgment transitions.
Additionally, we used paired-sample $t$ tests to evaluate cross-phase differences in participant-level FPR and FNR. We report the means and standard deviations for PhaseI and PhaseII, alongside the mean paired difference, its 95\% confidence interval, the $t$ statistic, the $p$ value, and Cohen’s $d_z$.


\subsubsection{Cross-Dataset Comparison}
To answer RQ3, we repeated the accuracy, binary-response, correctness-transition, FPR, and FNR analyses for the DFDC evaluation. Because the same cohort of participants completed evaluations on both datasets, we compared the magnitude of their cross-phase accuracy changes using a paired-sample $t$-test. For each participant, we derived the accuracy change from PhaseI to PhaseII for FakeAVCeleb and DFDC, respectively, and tested for a significant difference between these dataset-specific change scores. We report the mean difference, the 95\% confidence interval of the difference, the $t$ statistic, the $p$ value, and Cohen's $d_z$.
Because all participants completed the FakeAVCeleb evaluation before the DFDC evaluation, this cross-dataset comparison cannot distinguish between dataset characteristics and practice, transfer, fatigue, or other order effects. Therefore, the comparison results should be interpreted as descriptive comparisons within participants under a fixed evaluation order, rather than causal dataset effects.


\subsubsection{Confidence Analysis}
One-sample $t$-tests assessed whether the mean signed confidence-accuracy gap (Eq.~(\ref{eq:signed_confidence_gap})) differed from zero. Pearson and Spearman correlation analyses examined the association between confidence and pooled accuracy (Eq.~(\ref{eq:pooled_accuracy})), while paired-sample $t$-tests compared confidence and signed confidence-accuracy gaps between the two dataset evaluations.

\begin{table*}[t]
\centering
\caption{Participant performance by dataset and phase. Performance is presented as mean$\pm$standard deviation for PhaseI and PhaseII. Differences are calculated subtracting PhaseI from PhaseII and are expressed in percentage points.}
\label{tab:participant_performance}
\renewcommand{\arraystretch}{1.15}
\setlength{\tabcolsep}{5pt}
\begin{tabular}{llcccccc}
\hline
\textbf{Dataset} &
\textbf{Metric} &
\textbf{PhaseI} &
\textbf{PhaseII} &
\textbf{Difference [95\% CI]} &
\textbf{$t(109)$} &
\textbf{$p$} &
\textbf{$d_z$} \\
\hline
FakeAVCeleb
& Accuracy
& 63.30$\pm$12.99
& 67.98$\pm$13.67
& 4.68 [3.03, 6.33]
& 5.61
& $<.001$
& 0.54 \\

& FPR
& 24.45$\pm$19.90
& 25.23$\pm$20.85
& 0.77 [$-2.28$, 3.83]
& 0.50
& .617
& 0.05 \\

& FNR
& 48.95$\pm$24.01
& 38.82$\pm$24.96
& $-10.14$ [$-13.75$, $-6.53$]
& $-5.57$
& $<.001$
& $-0.53$ \\
\hline
DFDC
& Accuracy
& 59.52$\pm$11.84
& 68.16$\pm$15.39
& 8.64 [6.31, 10.96]
& 7.37
& $<.001$
& 0.70 \\

& FPR
& 27.55$\pm$18.86
& 25.09$\pm$20.73
& $-2.45$ [$-5.86$, 0.95]
& $-1.43$
& .156
& $-0.14$ \\

& FNR
& 53.41$\pm$23.74
& 38.59$\pm$27.53
& $-14.82$ [$-19.18$, $-10.45$]
& $-6.73$
& $<.001$
& $-0.64$ \\
\hline
\end{tabular}
\end{table*}

\section{Results}
\label{sec:results}


\vspace{-2mm}
\begin{table}[b]
\centering
\caption{Response transition between PhaseI and PhaseII. Each cell reports the count and percentage of a specific case out of 4,400 participant-video pairs in each dataset.}
\label{tab:binary_response_transitions}
\renewcommand{\arraystretch}{1.15}
\setlength{\tabcolsep}{5pt}
\begin{tabular}{llcc}
\hline
\textbf{Dataset} &
\textbf{PhaseI response} &
\multicolumn{2}{c}{\textbf{PhaseII response}} \\
\cline{3-4}
& & \textbf{Fake} & \textbf{Real} \\
\hline
FakeAVCeleb
& Fake
& 1,175 (26.70\%)
& 486 (11.05\%) \\

& Real
& 726 (16.50\%)
& 2,013 (45.75\%) \\
\hline
DFDC
& Fake
& 1,043 (23.70\%)
& 588 (13.36\%) \\

& Real
& 860 (19.55\%)
& 1,909 (43.39\%) \\
\hline
\end{tabular}
\end{table}

\subsection{RQ1: Accuracy Change Across Reassessment}
\label{subsec:rq1_results}

Table~\ref{tab:participant_performance} summarizes participant performance by dataset and phase. For FakeAVCeleb, the mean accuracy of participants improved by 4.68 percentage points ($SD=8.75$, 95\% CI [3.03, 6.33]) from 63.30\% ($SD=12.99$) in PhaseI to 67.98\% ($SD=13.67$) in PhaseII. Paired-sample $t$ tests indicated that this improvement was statistically significant ($t(109)=5.61$, $p<.001$) and the within-participant effect size was moderate ($d_z=0.54$).

The complementary judgment-level logistic regression analysis yielded the same directional results. With sandwich standard errors clustered by both participant and video, the estimated phase coefficient was positive and statistically significant ($\beta=0.208$, $SE=0.040$, $z=5.23$, $p<.001$). The corresponding odds ratio was 1.23 (95\% CI [1.14, 1.33]), indicating that the probability of making a correct judgment in PhaseII was approximately 23\% higher than in PhaseI.

Therefore, both participant-level and judgment-level analyses supported the same conclusion: the accuracy of reassessments was higher. However, practical improvement was limited, and the mean accuracy in PhaseII remained below 70\%, with approximately one-third of judgments still being incorrect.

Furthermore, for FakeAVCeleb, as shown in Table~\ref{tab:participant_performance}, the increase in accuracy from PhaseI to PhaseII was mainly attributed to the reduction in false-negative errors. The mean false-negative rate decreased from 48.95\% ($SD=24.01$) in PhaseI to 38.82\% ($SD=24.96$) in PhaseII. The mean difference was $-10.14$ percentage points (95\% CI [$-13.75$, $-6.53$]) and was statistically significant ($t(109)=-5.57$, $p<.001$, $d_z=-0.53$).
In contrast, the mean false-positive rate changed from 24.45\% ($SD=19.90$) in PhaseI to 25.23\% ($SD=20.85$) in PhaseII. The mean difference was 0.77 percentage points (95\% CI [$-2.28$, 3.83]) and was not statistically significant ($t(109)=0.50$, $p=.617$, $d_z=0.05$).

\subsection{RQ2: Judgment Stability, Correctness Transitions, and Error Types}
\label{subsec:rq2_results}

As shown in Table~\ref{tab:binary_response_transitions}, of 4,400 participant-video pairs from FakeAVCeleb, 1,175 (26.70\%) maintained the fake response (from fake to fake), 486 (11.05\%) changed from fake to real, 726 (16.50\%) changed from real to fake, and 2013 (45.75\%) maintained the real response (from real to real). 
Furthermore, participants maintained the same real/fake response in 72.45\% of the pairs, while their responses reversed in 27.55\% of the pairs.
Although more than a quarter of the responses reversed, most responses remained stable across phases.

The correctness-transition results are shown in Table~\ref{tab:correctness_transitions}. For FakeAVCeleb, of 4,400 participant-video pairs, 2,282 (51.86\%) were correct in both phases, while 906 (20.59\%) were incorrect in both phases. Furthermore, 709 (16.11\%) changed from incorrect in PhaseI to correct in PhaseII, while 503 (11.43\%) changed from correct to incorrect. Of the 1,615 pairs that were incorrect in PhaseI, 709 (43.90\%) were corrected in PhaseII.
Of the 2,785 pairs that were correct in PhaseI, 503 (18.06\%) became incorrect in PhaseII.
These results show that error correction transitions were more frequent than new-error formation transitions, and the continuity-corrected McNemar test showed a significant difference ($\chi^2(1)=34.67$, $p<.001$). Since the participant-video pairs were clustered within both participants and videos, this test result should be considered as supplementary to the descriptive transition results, rather than evidence from 4,400 independent observations.

\vspace{-4mm}
\begin{table}[b]
\centering
\caption{Correctness transition between PhaseI and PhaseII. Percentages are calculated with 4,400 participant-video pairs per dataset as the denominator.}
\label{tab:correctness_transitions}
\renewcommand{\arraystretch}{1.15}
\setlength{\tabcolsep}{4pt}
\begin{tabular}{lrrrr}
\hline
&
\multicolumn{2}{c}{\textbf{FakeAVCeleb}} &
\multicolumn{2}{c}{\textbf{DFDC}} \\
\cline{2-3}
\cline{4-5}
\textbf{Transition} &
\textbf{Count} &
\textbf{\%} &
\textbf{Count} &
\textbf{\%} \\
\hline
Correct $\rightarrow$ Correct
& 2,282 & 51.86
& 2,085 & 47.39 \\

Incorrect $\rightarrow$ Incorrect
& 906 & 20.59
& 867 & 19.70 \\

Incorrect $\rightarrow$ Correct
& 709 & 16.11
& 914 & 20.77 \\

Correct $\rightarrow$ Incorrect
& 503 & 11.43
& 534 & 12.14 \\
\hline
\textbf{Total}
& \textbf{4,400} & \textbf{100.00}
& \textbf{4,400} & \textbf{100.00} \\
\hline
\end{tabular}
\end{table}

\subsection{RQ3: Cross-Dataset Consistency}
\label{subsec:rq3_results}

\subsubsection{Participant Performance on DFDC}
As shown in Table~\ref{tab:participant_performance}, for DFDC, the mean accuracy of participants improved by 8.64 percentage points ($SD=12.29$, 95\% CI [6.31, 10.96]) from 59.52\% ($SD=11.84$) in PhaseI to 68.16\% ($SD=15.39$) in PhaseII. Paired-sample t tests indicated that the improvement was statistically significant ($t(109)=7.37$, $p<.001$) and the within-participant effect size was moderate ($d_z=0.70$).

The complementary judgment-level logistic regression analysis yielded the same conclusion. With sandwich
standard errors clustered by both participant and video, the phase coefficient was positive and statistically significant ($\beta=0.375$, $SE=0.065$, $z=5.78$, $p<.001$). The corresponding odds ratio was 1.46 (95\% CI [1.28, 1.65]), indicating that the probability of making a correct judgment in PhaseII was approximately 46\% higher than in PhaseI.

A similar trend was observed in the FakeAVCeleb dataset as in the DFDC dataset, the increase in accuracy from PhaseI to PhaseII was mainly attributed to fewer fake videos being judged as real. The mean false-negative rate decreased from 53.41\% ($SD=23.74$) in PhaseI to 38.59\% ($SD=27.53$) in PhaseII. The mean difference was $-14.82$ percentage points (95\% CI [$-19.18$, $-10.45$]) and was statistically significant ($t(109)=-6.73$, $p<.001$, $d_z=-0.64$). In contrast, the mean false-positive rate decreased from 27.55\% ($SD=18.86$) in PhaseI to 25.09\% ($SD=20.73$) in PhaseII. The mean difference was $-2.45$ percentage points (95\% CI [$-5.86$, 0.95]) and was not statistically significant ($t(109)=-1.43$, $p=.156$, $d_z=-0.14$). 

\subsubsection{Judgment Transition on DFDC}
As shown in Table~\ref{tab:binary_response_transitions}, of 4,400 participant-video pairs from DFDC, 1,043 (23.70\%) maintained the fake response (from fake to fake), 588 (13.36\%) changed from fake to real, 860 (19.55\%) changed from real to fake, and 1,909 (43.39\%) maintained the real response (from real to real). Furthermore, participants maintained the same response in 67.09\% of the pairs and reversed their response in 32.91\% of the pairs.

As shown in Table~\ref{tab:correctness_transitions}, of 4,400 participant-video pairs, 2,085 (47.39\%) were correct in both phases, 867 (19.70\%) were incorrect in both phases, 914 (20.77\%) changed from incorrect in PhaseI to correct in PhaseII, and 534 (12.14\%) changed from correct in PhaseI to incorrect in PhaseII.
Of the 1,781 pairs that were incorrect in PhaseI, 914 (51.32\%) were corrected in PhaseII.
Of the 2,619 pairs that were correct in PhaseI, 534 (20.39\%) became incorrect in PhaseII.
As in FakeAVCeleb, error-correction transitions were more frequent than new-error-formation transitions, and the continuity-corrected McNemar test showed a significant difference ($\chi^2(1)=99.20$, $p<.001$). As in the analysis on FakeAVCeleb, this test result should
be considered as supplementary to the descriptive transition results, rather than evidence from 4,400 independent observations, because the participant-video pairs were clustered within both participants and videos.

\subsubsection{Comparison between FakeAVCeleb and DFDC}
As shown in Table~\ref{tab:participant_performance}, the improvement in accuracy from PhaseI to PhaseII on DFDC was 3.96 percentage points (95\% CI [0.82, 7.09]) higher than that on FakeAVCeleb, and the difference is statistically significant ($t(109)=2.50$, $p=.014$, $d_z=0.24$). Because all participants completed the evaluation on FakeAVCeleb before the evaluation on DFDC, this difference may reflect a combination of dataset characteristics, practice, transfer, fatigue, or other order effects. Therefore, the test result should be interpreted as descriptive comparisons within participants under a fixed evaluation order, rather than causal dataset effects.

Except for the difference mentioned above, the results in Tables~\ref{tab:participant_performance} through~\ref{tab:correctness_transitions} show that the participants' performance patterns were consistent across the two datasets.

\subsection{Confidence-Accuracy Analysis}
\label{subsec:confidence_results}

For FakeAVCeleb, the mean confidence level of participants was 77.63\% ($SD=15.61$), while the mean pooled accuracy was 65.64\% ($SD=12.60$). The mean signed confidence-accuracy gap was 11.99 percentage points ($SD=16.18$, 95\% CI [8.93, 15.05]), which was significantly greater than zero ($t(109)=7.77$, $p<.001$). The mean absolute confidence-accuracy gap was 15.92 percentage points ($SD=12.29$). Participants' confidence and accuracy was significantly positively correlated (Pearson correlation coefficient $r=.358$, $p<.001$; Spearman's rank correlation coefficient $\rho=.348$, $p<.001$).

For DFDC, the mean confidence was 76.66\% ($SD=18.82$), while the mean pooled accuracy was 63.84\% ($SD=12.28$). The mean signed confidence-accuracy gap was 12.82 percentage points ($SD=21.57$, 95\% CI [8.75, 16.90]), and was significantly greater than zero ($t(109)=6.24$, $p<.001$). The mean absolute confidence-accuracy gap was 21.15 percentage points ($SD=13.41$). There was no significant correlation between participants' confidence and accuracy (Pearson's $r=.086$, $p=.373$; Spearman's $\rho=.074$, $p=.441$).

The difference in mean confidence between the two evaluations on FakeAVCeleb and DFDC was $-0.96$ percentage points (DFDC minus FakeAVCeleb), which was not significant ($t(109)=-0.42$, $p=.676$). The difference in the signed confidence-accuracy gap was $0.83$ percentage points, also not significant ($t(109)=0.35$, $p=.725$).

These analyses demonstrate a consistent group-level confidence-accuracy mismatch in both evaluations. Because confidence was measured only after both phases, they did not establish whether confidence or confidence calibration changed from PhaseI to PhaseII.

\section{Discussion}
\label{sec:discussion}

\subsection{Reassessment Improved Accuracy, but the Benefit Was Limited}

For both the FakeAVCeleb and DFDC evaluations, the accuracy of PhaseII was higher than that of PhaseI, indicating that reassessment combined with corrective feedback is associated with improved performance. The consistent trend between the two evaluations suggests that this effect is not limited to a single video set. However, the accuracy of PhaseII remained below 70\%, indicating that prior exposure and item-specific feedback do not enable humans to make reliable judgments without assistance.

Several mechanisms could explain this improvement. Participants might have remembered previously presented labels, identified specific features of the stimuli, reconsidered their response strategy, or paid closer attention to possible manipulation cues. However, given the combination of repeated presentation and corrective feedback and the use of the same videos in both phases, improvement cannot be attributed solely to repetition, feedback, perceptual learning, familiarity, or label memory. It is important to note that this study did not explore whether participants developed a transferable ability to identify deepfake content in previously unseen videos.

The improvement was greater for DFDC than for FakeAVCeleb, but this difference should be interpreted with caution. All participants completed the FakeAVCeleb evaluation first, followed by the DFDC evaluation; therefore, the dataset characteristics were confounded with practice, transfer, fatigue, and other order effects. The most defensible cross-dataset conclusion is therefore that the improvement in PhaseII was validated under the fixed evaluation order.

\subsection{Reassessment Corrected Errors but Also Introduced New Ones}

The transition analysis in Table~\ref{tab:binary_response_transitions} shows that reassessment is not simply a replication of participants' initial decisions. Although most of the responses remained unchanged, more than one-quarter of the judgments on FakeAVCeleb and approximately one-third of the judgments on DFDC were reversed. Beneficial reversals were more common than detrimental reversals; i.e., participants corrected more initial errors than replaced correct answers with incorrect ones.

However, the transition patterns also reveal considerable instability. Approximately one-fifth of judgments remained incorrect in both phases, and some initially correct judgments became incorrect during reassessment. Therefore, despite the overall improvement in accuracy, the transition analysis revealed three distinct behavioral outcomes: successful error correction, persistent incorrect judgments throughout both phases, and the emergence of new errors during reassessment.

Participants may correct errors by recalling feedback or noticing previously ignored audiovisual information. Persistent errors may reflect failure to remember feedback, confusion with the displayed label, or persistent reliance on nondiagnostic cues. New errors may result from second-guessing, imperfect source memory, or a shift in attention to misleading features. These explanations remain hypothetical, as this study did not directly measure identification memory, familiarity, attention, or response strategies.

The coexistence of error correction and new-error formation indicates that familiarity with a video does not guarantee an accurate recall of its authenticity. The findings are consistent with theories of perceptual learning and source monitoring, but they do not demonstrate a mere-exposure or illusory-truth effect because familiarity and perceived authenticity were neither independently manipulated nor measured.

\subsection{Accuracy Gains Primarily Reflected Fewer False Negatives}

The class-specific FPR and FNR analysis shows similar patterns on FakeAVCeleb and DFDC. FNR decreased in PhaseII compared to PhaseI, while FPR did not change significantly. Therefore, the overall improvement in accuracy in PhaseII compared to PhaseI was primarily attributed to fewer manipulated videos mistakenly identified as genuine.

The decrease in FNR is potentially beneficial, as mistaking manipulated content for genuine content can facilitate misinformation and fraud. However, the FNR in PhaseII remained high, indicating that many manipulated videos could still deceive participants even after prior exposure and corrective feedback. Furthermore, corrective feedback did not effectively reduce participants' tendency to label authentic videos as fake, an error that can undermine trust in legitimate audiovisual evidence.

The observed patterns may reflect improved discriminatory abilities, a greater inclination to answer ``fake,'' or both. Accuracy, FPR, and FNR cannot distinguish perceptual sensitivity from response bias. Therefore, future research should incorporate signal detection metrics such as sensitivity and response criteria.

\subsection{Confidence Exceeded Observed Performance}

Participants' confidence levels were higher than their accuracy levels in both evaluations. This mismatch suggests that participants tended to overestimate their performance, which was inconsistent with their actual judgments. Furthermore, we observed a correlation between confidence and performance on FakeAVCeleb, but not on DFDC. Therefore, we cannot infer whether participants with higher confidence levels also performed well.

These findings should not be interpreted as a complete evaluation of confidence calibration. Each participant provided their confidence level only after both phases of evaluations for each dataset, rather than after each individual judgment or each phase. Therefore, these data cannot determine whether confidence levels changed after feedback, whether participants gained appropriate confidence after correcting errors, or whether persistent and new errors were accompanied by greater certainty.

This mismatch remains relevant to practical validation. Participants who overestimate their detection abilities may be less willing to seek corroborating evidence or automated assistance. However, since this study did not measure information-seeking behavior or reliance on external validation tools, this inference still requires direct verification.

\subsection{Implications for Human-AI Verification}

The findings indicate that unassisted human judgment remains vulnerable when repeatedly exposed to audiovisual deepfakes. Although reassessment after item-specific feedback produced measurable improvement, persistent errors, judgment reversals, new errors, and mismatches between confidence and accuracy remained common. Therefore, human memory of previous exposure or feedback should not be considered a reliable verification mechanism.

These results motivate the development of verification systems that combine human judgment with automated assistance. When human judgment is flawed, the interface should not only provide binary corrective feedback but also prompt for further examination, such as highlighting potentially diagnostic evidence like facial artifacts, acoustic synthesis traces, temporal irregularities, or audiovisual mismatches. 

Such systems can also assist in confidence calibration by identifying high-confidence errors and suggesting additional verification when decisions have significant consequences. However, this study did not evaluate AI-assisted decision-making. Whether detector outputs, uncertainty displays, or explanatory evidence improve human performance must be established through controlled experiments that measure not only accuracy but also appropriate reliance, response bias, confidence calibration, and performance on previously unseen content.

\section{Limitations}
\label{sec:limitations}

Several limitations constrain the interpretation of this study. First, repeated exposure and corrective feedback were combined, and Phase II reused the same videos. Therefore, the observed improvement may reflect familiarity, item-specific memory, response strategy changes, or feedback use rather than a transferable enhancement in deepfake detection ability. Memory, attention, familiarity, and perceived truth were not directly measured.

Second, all participants completed the FakeAVCeleb evaluation before the DFDC evaluation, meaning that dataset differences may be influenced by order effects, practice, transfer, or fatigue. In addition, the use of 40 videos per dataset limits generalization across speakers, manipulation methods, recording conditions, languages, and emerging deepfake systems.

Third, confidence was measured only once after both phases, allowing assessment of group-level confidence-accuracy correspondence but not changes in calibration or confidence following feedback.

Finally, the sample consisted of English-speaking adults recruited through Amazon Mechanical Turk, and the study examined only unassisted human judgments. Thus, the findings may not generalize to other populations or to decisions made with automated assistance.

\section{Future Work}
\label{sec:future_work}

Future research should disentangle the effects of repeated exposure, corrective feedback, and item-specific memory using factorial designs that incorporate novel stimuli, feedback conditions, longer retention intervals, and counterbalanced dataset order.

Larger and more diverse stimulus sets, including recent generation methods, multiple manipulation modalities, compression levels, recording conditions, and languages, should be used, with transparent stimulus selection to improve reproducibility.

Judgment-level confidence, signal detection measures, and hierarchical models should be incorporated to evaluate confidence calibration, distinguish perceptual sensitivity from response bias, and account for variability across participants and videos.

Finally, future studies should investigate human-AI collaboration by evaluating detector probabilities, uncertainty displays, and explanatory feedback, while assessing accuracy, response time, confidence calibration, appropriate reliance on automation, and generalization to unseen deepfakes.

\section{Conclusion}
\label{sec:conclusion}

This study examined how human judgments of audiovisual deepfakes changed when participants reassessed the same videos after receiving item-specific corrective feedback. Mean accuracy increased from 63.30\% to 67.98\% on FakeAVCeleb and from 59.52\% to 68.16\% on DFDC, demonstrating consistent cross-phase improvement across both datasets. However, because the datasets were evaluated in a fixed order, differences in the magnitude of improvement cannot be attributed solely to dataset characteristics.

Although reassessment corrected more errors than it introduced, persistent errors remained (20.59\% for FakeAVCeleb and 19.70\% for DFDC), and some initially correct judgments became incorrect (18.06\% and 20.39\%, respectively). The accuracy gains were primarily driven by a reduction in false negatives, while false-positive rates remained largely unchanged. Participants also reported confidence levels that exceeded their observed accuracy.

Overall, reassessment with corrective feedback improved deepfake detection accuracy but did not eliminate persistent or newly introduced errors. Because repeated exposure and corrective feedback were combined using the same stimuli, the observed improvements cannot be attributed to any single mechanism or assumed to generalize to previously unseen content. Further research should isolate these effects and evaluate human-AI collaboration for deepfake verification.

\bibliographystyle{IEEEtran}
\bibliography{references}

\newpage

\vfill

\end{document}